\documentclass[12pt]{article}
\usepackage{epsfig}
\usepackage{amssymb}
\usepackage{amsmath}
\usepackage{amsfonts}
\usepackage{amsthm}
\usepackage{graphicx}
\usepackage{mathrsfs}
\usepackage[dvips]{color}
\usepackage{multirow}

\newcommand{\bSigma}{\boldsymbol{\Sigma}}

\newcommand{\bnabla}{\boldsymbol{\nabla}}

\newcommand{\R}{\mathbb{R}}
\newcommand{\C}{\mathbb{C}}
\newcommand{\Z}{\mathbb{Z}}
\newcommand{\N}{\mathbb{N}}

\newcommand{\fz}{\mathfrak{z}}

\newcommand{\fK}{\mathfrak{K}}

\newcommand{\bfe}{\mathbf{e}}

\newcommand{\bk}{\mathbf{k}}

\newcommand{\bp}{{\mathbf{p}}}
\newcommand{\bq}{{\mathbf{q}}}

\newcommand{\bx}{\mathbf{x}}
\newcommand{\bu}{\mathbf{u}}

\newcommand{\bcE}{{\boldsymbol{\cE}}}
\newcommand{\bF}{\mathbf{F}}

\newcommand{\bI}{\mathbf{I}}

\newcommand{\bQ}{\mathbf{Q}}

\newcommand{\bV}{\mathbf{V}}

\newcommand{\cH}{\mathcal{H}}
\newcommand{\cE}{\mathcal{E}}

\newcommand{\be}{\begin{equation}}
\newcommand{\ee}{\end{equation}}
\newcommand{\bea}{\begin{eqnarray}}
\newcommand{\eea}{\end{eqnarray}}
\newcommand{\nn}{\nonumber}
\newcommand{\kt}{\rangle}
\newcommand{\br}{\langle}

\newcommand{\ed}{\end{document}}

\newcommand{\bi}{\begin{itemize}}
\newcommand{\ei}{\end{itemize}}

\newcommand{\bce}{\begin{center}}
\newcommand{\ece}{\end{center}}

\newcommand{\sF}{\mathscr{F}}

\newcommand{\bPsi}{{\boldsymbol{\Psi}}}
\newcommand{\bPhi}{{\boldsymbol{\Phi}}}

\newcommand{\bcH}{{\boldsymbol{\cH}}}

\newcommand{\bvarepsilon}{{{\mbox{$\boldsymbol{\varepsilon}$}}}}
\newcommand{\bmu}{{{\mbox{$\boldsymbol{\mu}$}}}}

\newcommand{\bzero}{{\boldsymbol{0}}}

\newcommand{\for}{{\mbox{\rm for}}}

\newcommand{\bfeta}{{\boldsymbol{\eta}}}

\begin{document}

\title{Can $N$-th Order Born Approximation Be Exact?}

\author{Farhang Loran\thanks{E-mail address: loran@iut.ac.ir}
~and Ali~Mostafazadeh\thanks{Corresponding author, E-mail address:
amostafazadeh@ku.edu.tr}\\[6pt]
$^*$Department of Physics, Isfahan University of Technology, \\ Isfahan 84156-83111, Iran\\[6pt]
$^\dagger$Departments of Mathematics and Physics, Ko\c{c} University,\\  34450 Sar{\i}yer,
Istanbul, T\"urkiye}

\date{ }
\maketitle

\begin{abstract}
For the scattering of scalar waves in two and three dimensions and electromagnetic  waves in three dimensions, we identify a condition on the scattering interaction under which the $N$-th order Born approximation gives the exact solution of the scattering problem for some $N\geq 1$.
\end{abstract}

\section{Introduction}

Born's formulation of quantum scattering theory \cite{Born-1926} is among the most important achievements of the 20th century theoretical physics. It provided the foundations up on which modern scattering theory is build. It also offered a perturbative method of calculating scattering amplitudes known as the $N$-th Born approximation, where $N$ signifies the order of the perturbation. This approximation scheme was subsequently generalized to the scattering of electromagnetic waves \cite{born-wolf,TKD} and found a wide range of applications \cite{hofstadter,Breuer-1981,koshino,Hunter-2006,Bennett-2014,Bereza-2017,van der Sijs}. This has made Born approximation into an indispensable part of standard textbooks and monographs on scattering theory \cite{newton,taylor}. Yet the natural question of finding conditions under which it gives the exact solution of a scattering problem was left unanswered for over nine decades. The purpose of the present article, is to offer such a condition for the scattering of scalar waves in two and three dimensions and electromagnetic waves in three dimensions.

It is well-known that the scattering amplitude for the Coulomb potential is proportional to its coupling constant. This is sometimes taken as a sign that the first Born approximation is exact for the Coulomb potential. One must however note that this is a long-range potential whose scattering problem cannot be properly handled by standard Born approximation. Indeed, a careful examination shows that the first Born approximation fails to produce the correct expression for the phase of the scattering amplitude and the second-order term in the Born series for the scattering amplitude blows up \cite{newton}. Therefore strictly speaking the Born approximation does not give the exact solution of the scattering problem for the Coulomb potential. 

In Refs.~\cite{pra-2019,pra-2021} we identify a class of short-range potentials in two and three dimensions whose scattering problems are exactly solvable by the first Born approximation if the wavenumber of the incident wave does not exceed a preassigned characteristic value. In Ref.~\cite{ptep-2024} we generalize this result to the scattering of electromagnetic (EM) waves by a general stationary linear medium. This allows for identifying optical media displaying broadband directional invisibility \cite{apl-2023}. 

The results of Refs.~\cite{pra-2019,pra-2021,ptep-2024} make extensive use of a recently developed dynamical formulation of stationary scattering \cite{pra-2021, ap-2014,pra-2016,pra-2023a} which relies on a fundamental notion of transfer matrix. In one dimension this coincides with the standard transfer matrix used to conduct scattering calculations since the 1940's \cite{jones-1941,yeh,griffiths-2001,sanchez-soto,tjp-2020}. In higher dimensions it is an integral operator that similarly to the S matrix contains the information about the scattering properties of the interaction. In the present article, we use the standard formulation of stationary scattering that is based on the Lippmann-Schwinger equation \cite{Sakurai} to obtain a sufficient condition for the exactness of the $N$-th order Born approximation in potential scattering in two and three dimensions and EM scattering in three dimensions. 

{The organization of this article is as follows. In Sec.~2 we offer a concise review of potential scattering for scalar waves in two and three dimensions, and state and prove a theorem on the exactness of the $N$-th order Born approximation. In Sec.~3 we generalize the results of Sec.~2 to the scattering of electromagnetic waves propagating in a general linear stationary medium in three dimensions. In Sec.~4, we present our concluding remarks.}

In what follows we use $\N$, $\R$, $\R^+$, and $\Z^+$ respectively for the sets of natural numbers (nonnegative integers), real numbers, positive real numbers, and positive integers.

\section{Potential scattering}

Consider a scattering process in two or three dimensions in which an incident plane wave with wave vector $\bk$ scatters off a possibly complex-valued potential $v:\mathbb{R}^d\to\mathbb{C}$ where $d\in\{2,3\}$ and $|v(\bx)|$ tends to zero faster than $|\bx|^{-(d+1)/2}$, \cite{Yafaev}. The scattering data collected by a detector that is located at a point $\bx$ far away from the scattering region is encoded in the scattering amplitude $f$ of the potential. The latter is defined in terms of the asymptotic expression,
	\be
      	\psi(\bx)\to \frac{1}{(2\pi)^{\frac{d}{2}}} \left[e^{i\bk\cdot\bx}+
	f(\bk',\bk)\,\frac{e^{ik|\bx|}}{|\bx|^\frac{d-1}{2}}\right]
	\quad \for\quad|\bx|\to\infty,
    	\label{scattering-amplitude}
    	\ee
for the scattering solutions of the stationary Schr\"odinger equation,
	\be
    	\left[-\nabla^2+v(\bx) \right]\psi(\bx)=k^2\psi(\bx),
    	\label{sch-eq}
    	\ee
where $\bk':=k\frac{\bx}{|\bx|}$ is the scattered wave vector, and $k:=|\bk|$ is the incident wavenumber. More precisely, $\psi(\bx)$ is the position wave function for  the solution $|\psi\kt$ of the Lippmann-Schwinger equation,
    \be
    |\psi\kt= |\bk\kt+ 
    G(\widehat \bp)v(\widehat\bx)|\psi\kt,
   \label{LP-ket}
   \ee
where $\br\bx|\bk\kt:=(2\pi)^{-d/2}e^{i\bk\cdot\bx}$, $\hat\bp$ and $\hat\bx$ are respectively the standard momentum and position operators in $d$ dimensions, and $G(\bp):=(k^2-{\bp}^2+i\epsilon)^{-1}$ with $\epsilon\to 0^+$, \cite{Sakurai}. It is important to note that if $v$ is a complex potential, the scattering amplitude may blow up at certain (real and positive) values of $k$. In this case $k^2$ is called a spectral singularity \cite{prl-2009} and the system amplifies background noise to emit purely out-going radiation, a phenomenon which is realized in every laser \cite{pra-2013-2,pra-2013-4,jaosb-2020}. Throughout this article we confine our attention to generic situations where $k^2$ is not a spectral singularity.

In the momentum representation, the Lippmann-Schwinger equation \eqref{LP-ket} reads
    	\be
     	\br\bp|\psi\kt= \br\bp|\bk\kt+ \frac{G(\bp)}{(2\pi)^d}\int_{\mathbb{R}^d} d\bq\, \widetilde v(\bp-\bq) \br\bq|\psi\kt,
    	\label{LP-p-rep}
    	\ee
where $\widetilde{v}:\mathbb{R}^d\to\mathbb{C}$ denotes the Fourier transform of $v$, i.e., $\widetilde{v}(\bp):=\int_{\R^d}d^dx\:e^{-i\bp\cdot\bx}\:v(\bx)$. Repeated use of  (\ref{LP-p-rep}) in its right-hand side leads to the Born series,
    	\be
     	\br\bp|\psi\kt=\sum_{n=0}^\infty\br\bp|\psi^{(n)}\kt,
    	\label{Born}
    	\ee
where $\br\bp|\psi^{(0)}\kt:=\br\bp|\bk\kt$ and
	\be
     	\br\bp|\psi^{(n+1)}\kt:=\frac{G(\bp)}{(2\pi)^d}
	\int_{\mathbb{R}^d} d\bq\, \widetilde v(\bp-\bq) 	
	\br\bq|\psi^{(n)}\kt\quad\for\quad n\geq 0.
    	\label{Born-iteration}
    	\ee
The $N$-th order Born approximation is the approximation in which one ignores the terms $\br\bp|\psi^{(n)}\kt$ with $n>N$ in (\ref{Born}), i.e.,
	\be
     	\br\bp|\psi\kt\approx\sum_{n=0}^N\br\bp|\psi^{(n)}\kt.
    	\nn
    	\ee

Because, for $|\bx|\to\infty$, $v(\bx)$ tends to zero faster than $|\bx|^{-(d+1)/2}$, solutions of the Schr\"odinger equation (\ref{sch-eq}) and consequently the Lippmann-Schwinger equation (\ref{LP-ket}) in the position representation tend to those of the Helmholtz equation $(\nabla^2+k^2)\psi(\bx)=0$ for $|\bx|\to\infty$. The latter are superpositions of the solutions of the form $e^{i\bp\cdot\bx}$ where $\bp\in\R^d$ and $|\bp|=k$. Now, suppose that $\bu\in\R^d$ is an arbitrary unit vector and for each $\bq\in\R^d$,
	\begin{align}
	&q_{\parallel}:=\bu\cdot\bq,
	&&\bq_\perp:=\bq-q_{\parallel}\bu.
	\label{notation}
	\end{align}
Then we can identify $\bq$ with $(q_\parallel,\bq_\perp)$ and express the condition $|\bp|=k$ in the form $\bp_\perp^2=k^2-p_\parallel^2$. According to this equation, if $|p_\parallel|>k$, there must be a unit vector $\bu'$ orthogonal to $\bu$ such that $\bu'\cdot\bp_\perp$ is imaginary. This identifies $e^{i\bp\cdot\bx}$ with an exponentially decaying or growing function for $\bx=t\bu'$ as $|t|\to\infty$. The solutions of the Lippmann-Schwinger equation (\ref{LP-ket}) in the position representation are bounded functions. Therefore the exponentially growing solutions are to be discarded. The exponentially decaying (evanescent) solutions are admissible, but for $|\bx|\to\infty$  they tend to zero much faster than the term proportional to $f(\bk',\bk)$ in (\ref{scattering-amplitude}). Because $\bu$ is an arbitrary unit vector, this argument implies that the asymptotic expression (\ref{scattering-amplitude}) for the solution of the Lippmann-Schwinger equation and consequently the scattering amplitude are determined by $\br\bp|\psi\kt$ with $|\bp|=k$. In particular,
	\be
	f(\bk',\bk)=0~~\mbox{if~ $\br\bp|\psi\kt=\br\bp|\bk\kt$ for $|\bp|=k$.}
	\label{condi-1}
	\ee
Furthermore, the $N$-th order Born approximation yields the exact solution of the scattering problem provided that 
	\be
	\br\bp|\psi\kt=\sum_{n=0}^{N}\br\bp|\psi^{(n)}\kt~~~~\for~~~~|\bp|=k.
	\label{exact}
	\ee
	
The following theorem offers a sufficient condition for (\ref{exact}).
	\begin{itemize}
	\item[]{\bf Theorem 1}: Consider the scattering of an incident plane wave by a short-range potential $v:\R^d\to\C$ that tends to zero faster than $|\bx|^{-(d+1)/2}$ as $|\bx|\to\infty$, where $d\in\{2,3\}$. Suppose that there are $\alpha\in\R^+$ and $\bu\in\R^d$ such that $|\bu|=1$ and
	\be
	\widetilde v(\bp)=0\quad\for\quad \bu\cdot\bp<\alpha.
	\label{potential}
	\ee
Let $k$ denote the wavenumber for the incident wave, and $N:=\lfloor 2k/\alpha\rfloor$, i.e., $N$ is the largest integer that is not larger than $2k/\alpha$, so that $k<\frac{1}{2}(N+1)\alpha$. Then the $N$-th order Born approximation provides the exact solution of the scattering problem. In particular, $v$ does not scatter incident waves with $k<\alpha/2$, and the first Born approximation is exact for $k<\alpha$.
	\end{itemize}
Before giving a proof of this Theorem we present some preliminary results. 
We begin by expressing (\ref{potential}) in the form
	\be
	\widetilde v(q_\parallel,\bq_\perp)=\theta(q_\parallel-\alpha)\widetilde v(q_\parallel,\bq_\perp),
	\label{potential-2}
	\ee
where $\theta$ is the step function, 
	\[\theta(x):=\left\{\begin{array}{ccc} 
	0&\for& x< 0,\\	
	1&\for& x\geq 0.
	\end{array}\right.\] 
Because $\widetilde v$ is a continuous function,  (\ref{potential-2}) implies $\widetilde v(\alpha,\bq_\perp)=0$. Next, we present a simple consequence of (\ref{potential}).
	\begin{itemize}
	\item[]{\bf Lemma 1}: Let $\bu\in\R^d$ such that $|\bu|=1$, $\mu,\nu\in\R$, $\phi:\R^d\to\C$ be a function that has Fourier transform, and $v:\R^d\to\C$ be a short-range potential that decays to zero faster than $|\bx|^{-(d+1)/2}$ as $|\bx|\to\infty$. Suppose that $v$ and $\phi$ fulfill the following conditions.
	\bea
	&&\widetilde v(\bp)=0~~~\for~~~\bu\cdot\bp< \mu,
	\label{L01-1}\\
	&&\widetilde \phi(\bp)=0~~~\for~~~\bu\cdot\bp< \nu.
	\label{L01-2}
	\eea 
Then,
	\be
	\int_{\R^d}d\bq\:\widetilde v(\bp-\bq)\widetilde \phi(\bq)=\theta(p_\parallel-\mu-\nu)
	\int_{\R^{d-1}}\!\!d\bq_\perp 
	\int_{\nu}^{p_\parallel-\mu}\!\! dq_{\parallel}\:\widetilde v(\bp-\bq)\widetilde \phi(\bq).
	\label{L01-3}
	\ee
	\item[]Proof: The left-hand side of (\ref{L01-3}) has the form ${\rm LHS}:=\int_{\R^{d-1}}d\bq_\perp \int_{-\infty}^{\infty} dq_{\parallel}\:\widetilde v(\bp-\bq)\widetilde \phi(\bq)$. In view of (\ref{L01-1}) and (\ref{L01-2}), the integrand in the latter expression vanishes for $p_\parallel-q_\parallel<\mu$ and $q_\parallel<\nu$. This identifies LHS with the right-hand side of (\ref{L01-3}).~~$\square$
	\end{itemize}

	\begin{itemize}
	\item[]{\bf Lemma 2}: Let $\bu\in\R^d$ such that $|\bu|=1$, $v:\R^d\to\C$ be a short-range potential that decays to zero faster than $|\bx|^{-(d+1)/2}$ as $|\bx|\to\infty$ and fulfills
	\be
	\widetilde v(\bp)=0~~~\for~~~\bu\cdot\bp< 0,
	\label{L1-0}
	\ee 
and $\psi$ be the solution of the Lippmann-Schwinger (\ref{LP-p-rep}). Then $\br\bp|\psi\kt=0$ for $\bu\cdot\bp<-k$.
	\item[]Proof: First, we use induction on $n$ to prove that $\br\bp|\psi^{(n)}\kt=0$ for all $n\geq 0$ and $\bu\cdot\bp<-k$. Because for such a $\bp$, $|\bk|=k<-\bu\cdot\bp\leq|\bp|$, we have $\bp\neq\bk$. This implies $\br\bp|\psi^{(0)}\kt=\br\bp|\bk\kt=\delta(\bp-\bk)=0$. Next, let $m\in\N$ be such that $\br\bp|\psi^{(m)}\kt=0$ for $\bu\cdot\bp<-k$. In view of Lemma~1, the latter relation together with (\ref{notation}) and (\ref{L1-0}) imply
	\begin{align}
	\int_{\mathbb{R}^d} d\bq\, \widetilde v(\bp-\bq) 	
	\br\bq|\psi^{(m)}\kt&=\theta(p_\parallel+k)
	\int_{\R^{d-1}}\!\!d\bq_\perp \int_{-k}^{p_\parallel}\!\! dq_{\parallel}\:\widetilde v(\bp-\bq)
	\br \bq|\psi^{(m)}\kt\nn\\
	&=0~~~\for~~~p_\parallel<-k.
	\label{L1-1}
	\end{align}
Substituting this equation in (\ref{Born-iteration}) with $n$ changed to $m$, we find $\br\bp|\psi^{(m+1)}\kt=0$. This proves that $\br\bp|\psi^{(n)}\kt=0$ for $p_\parallel<-k$ and all $n\geq 0$. Therefore $\br\bp|\psi\kt=0$ for $p_\parallel<-k$.~~$\square$
\end{itemize}
For potentials $v$ fulfilling (\ref{potential}), we can use Eq.~(\ref{LP-p-rep})  
and Lemmas~1 and~2 to arrive at the following useful identity.
	\begin{align}
	\br \bp|\psi\kt&=\br\bp|\bk\kt+
	\frac{\theta(p_\parallel-\alpha+k)G(\bp)}{(2\pi)^d}
	\int_{\R^{d-1}}\!\!d\bq_\perp 
	\int_{-k}^{p_\parallel-\alpha}\!\! dq_{\parallel}\:\widetilde v(\bp-\bq)\br\bq|\psi\kt.
	\label{L2-1}
	\end{align}	
\begin{itemize}
	\item[]{\bf Lemma 3}: Let $\alpha$, $\bu$, $v$ be as in Theorem~1, $\psi$ be the solution of the Lippmann-Schwinger equation (\ref{LP-p-rep}), and $\beta$ be a non-negative real number. Suppose that $k<\frac{1}{2}\alpha(\beta+1)$. Then $\br\bp|\psi\kt=\br\bp|\bk\kt$ for $-k\leq p_\parallel\leq k-\alpha\beta$.
	\item[]Proof: for $p_\parallel\leq k-\alpha\beta$, $k<\frac{1}{2}\alpha(\beta+1)$ implies
$p_\parallel-\alpha+k\leq 2k-\alpha(\beta+1)< 0$. Therefore $\theta(p_\parallel-\alpha+k)=0$, and (\ref{LP-p-rep}) and (\ref{L2-1}) imply $\br\bp|\psi\kt=\br\bp|\bk\kt$.
	~~$\square$
\end{itemize}

\begin{itemize}
	\item[]{\bf Lemma 4}: Let $k$, $\alpha$, $\bu$, and $v$ be as in Theorem~1, for all $n\in\N$, $\br\bp|\psi^{(n)}\kt$ be the terms of the Born series (\ref{Born}), and $N\in\N$. 
Then for all $m\in\Z^+$, 
	\be
	\br\bp|\psi^{(m)}\kt=0~~~\for~~-k\leq p_\parallel\leq k-(N-m+1)\alpha.
	\label{L4-1}
	\ee
	\item[]Proof: We prove this lemma by induction on $m$. According to Lemma 3 with $\beta=N$, $\br\bp|\psi\kt=\br\bp|\bk\kt=\br\bp|\psi^{(0)}\kt$ for $-k\leq p_\parallel\leq k-\alpha N$. This shows that $\br\bp|\psi^{(n)}\kt=0$ for $n\geq 1$ and $-k\leq p_\parallel\leq k-\alpha N$. For $n=1$ this proves (\ref{L4-1}) for $m=1$. Now suppose that there is some $m_\star\in\Z^+$ such that (\ref{L4-1}) holds for $m=m_\star$. Then in view of Lemma~1, we have
	\begin{align}
	\int_{-\infty}^\infty d\bq\:\widetilde v(\bp-\bq)\br\bq|\psi^{(m_\star)}\kt=&\;
	\theta\big(p_\parallel-k+(N-m_\star)\alpha\big)\times\nn\\
	&\quad\int_{\R^{d-1}}\!\!d\bq\int_{k-(N-m_\star+1)\alpha}^{p_\parallel-\alpha}\!\!\!\!dq_\parallel\:
	\widetilde v(\bp-\bq)\br\bq|\psi^{(m_\star)}\kt.
	\label{L3-2}
	\end{align}
For $-k\leq p_\parallel\leq k-(N-m_\star)\alpha$, we have $p_\parallel+(N-m_\star)\alpha-k\leq 0$. Therefore either $\theta\big(p_\parallel+(N-m_\star)\alpha-k\big)=0$ or the boundaries of the definite integral on the right-hand side of (\ref{L3-2}) coincide. In both cases, the latter vanishes. Using this result and (\ref{Born-iteration}), we conclude that for this range of values of $p_\parallel$, $\br\bp|\psi^{(m_\star+1)}\kt=0$.~~$\square$
\end{itemize}
\begin{itemize}
\item[]{\bf Proof of Theorem~1}: Suppose that $|\bp|=k$. Then, for $m\geq N+1$, we have 
	\[-k\leq p_\parallel\leq k<k-(N-m+1)\alpha.\] 
By virtue of this relation,
we can use Lemma 4 to infer that whenever $m\geq N+1$, 
$\br\bp|\psi^{(m)}\kt=0$ for $|p_\parallel| \leq k$. Therefore (\ref{exact}) holds, i.e., the $N$-th order Born approximation provides the exact solution of the scattering problem. For $k<\frac{\alpha}{2}$, we can set $N=0$ and conclude that $\br\bp|\psi\kt=\br\bp|\psi^{(0)}\kt=\br\bp|\bk\kt$ for $|\bp|=k$. In light of (\ref{condi-1}), this shows that the potential does not scatter the incident wave. For $k<\alpha$, we can set $N=1$ and find that $\br\bp|\psi^{(m)}\kt=0$ for $m>1$ and $|\bp|=k$. Therefore, (\ref{exact}) holds for $N=1$, and the first Born approximation gives the exact solution of the scattering problem.~~$\square$
\end{itemize}

{We conclude this section by offering a method for constructing potentials that satisfy the hypothesis of of Theorem~1 in three dimensions.}

{Let us choose a coordinate system in which $\bu$ is along the positive $x$ axis. Then $p_\parallel=p_x$ and (\ref{potential}) reads, $\widetilde v(p_x,p_y,p_z)=0$ for  $p_x<\alpha$. It is easy to show that this condition holds, if there is a function $f:\R^3\to\C$ satisfying
	\be
	v(x,y,z)=e^{i\alpha x}\int_0^\infty d\fK\: e^{i\fK x}f(\fK,y, z).
	\label{potential-x2}
	\ee
Choosing $f$ in such a way that  the right-hand side of this relation has Fourier transform and tends to zero faster than $(x^2+y^2+z^2)^{-1}$ as $x^2+y^2+z^2\to\infty$, we can construct a variety of potentials fulfilling the hypothesis of Theorem~1. A specific family of examples are given by
	\begin{align}
	&v(x,y,z)=\frac{e^{i\alpha x}g(y,z)}{(1-\frac{ix}{a})^{m+1}},
	\label{v=specifc}
	\end{align}
where ,
	\begin{align}
	&g(y,z):=
	\left\{\begin{array}{cc}
	\fz &\for~~|y|\leq\frac{\ell_y}{2}~{\rm and}~|z|\leq \frac{\ell_z}{2},\\
	0 & {\rm otherwise},\end{array}\right.\nn
	\end{align}
$\fz$ is a real or complex parameter, $a$, $\ell_y$, and $\ell_z$ are positive real parameters, and $m$ is a positive integer. This corresponds to setting $f(\fK,y,z)=a (a\fK)^m e^{-a\fK} g(y,z)/m!$. As seen from (\ref{v=specifc}) the parameter $\alpha$ enters the expression for both real and imaginary parts of the potential.}
 {Notice that sums of the potentials of the form (\ref{v=specifc}) with different choices for $a$, $\ell_y$, and $\ell_z$ also satisfy the hypothesis of Theorem~1. These potentials may, in principle, be realized in terms of a carefully engineered permittivity profile with a rectangular cross section and regions of gain and loss as discussed in Ref.~\cite{ptep-2024,apl-2023}.}

\section{Electromagnetic scattering}

Consider the scattering of an incident time-harmonic plane electromagnetic (EM) wave by a general stationary linear medium in three dimensions. Suppose that the medium has no free charges. Then the interaction of the wave with the medium is characterized by 
	\begin{align}
	&\delta\hat\bvarepsilon:=\hat\bvarepsilon-\bI,
	&&\delta\hat\bmu:=\hat\bmu-\bI,
	\label{etas}
	\end{align}
where $\hat\bvarepsilon$ and $\hat\bmu$ are respectively the relative permittivity and permeability tensors of the medium, and $\bI$ is the $3\times 3$ identity matrix. 
In general $\delta\hat\bvarepsilon$ and $\delta\hat\bmu$  are complex $3\times 3$ matrix-valued functions of space. We denote their entries by $\delta\hat\varepsilon_{ij}$ and $\delta\hat\mu_{ij}$. 

We can express the electric and magnetic fields of the wave in the form, $\varepsilon_0^{-1/2}e^{-ikct}\bcE(\bx)$ and  $\mu_0^{-1/2}e^{-ikct}\bcH(\bx)$, where $\varepsilon_0$ are $\mu_0$ are respectively the permittivity and permeability of vacuum, $c:=(\varepsilon_0\mu_0)^{-1/2}$ is the speed of light in vacuum, $k$ is the wavenumber, and $\bcE$ and $\bcH$ are complex vector-valued functions of space which in view of Maxwell's equations \cite{jackson} satisfy
	\begin{align}
	&\bnabla\times\bcE(\bx)=ik\,\hat\bmu(\bx)\bcH(\bx),
	&&\bnabla\times\bcH(\bx)=-ik\,\hat\bvarepsilon(\bx)\bcE(\bx).
    	\label{Maxwell}
    	\end{align}
Taking the divergence of both sides of these equations and using (\ref{etas}), we have
	\begin{align}
	&\bnabla\cdot\bcE(\bx)=-\bnabla\cdot[\delta\hat\bvarepsilon(\bx)\,\bcE(\bx)], 
	&&\bnabla\cdot\bcH(\bx)=-\bnabla\cdot[\delta\hat\bmu(\bx)\,\bcH(\bx)].
    	\label{Maxwell-2}
    	\end{align}
	
Let us use $\C^{m\times n}$ and $\sF^{m\times n}$ to denote the vector spaces of $m\times n$ complex matrices and matrix-valued functions $\bF:\R^3\to\C^{m\times n}$, respectively. Then we can view $\bcE$ and $\bcH$ as elements of $\sF^{3\times 1}$ and introduce the $6$-component field,
	\begin{align}
	&\bPsi:=\left[\begin{array}{c}
	\bcE\\
	\bcH\end{array}\right]\in\sF^{\,6\times 1}.\nn
		\end{align}
Employing Dirac's bra-ket notation, we identify $\bcE$, $\bcH$, and $\bPsi$ respectively with $|\bcE\kt$, $|\bcH\kt$, and $|\bPsi\kt$, so that
	\begin{align}
	&\br\bx|\bcE\kt=\bcE(\bx), &&\br\bx|\bcH\kt=\bcH(\bx), 
	&&\br\bx|\bPsi\kt:=\bPsi(\bx)=\left[\begin{array}{c}
	\bcE(\bx)\\
	\bcH(\bx)\end{array}\right].\nn
	\end{align}
We also note that for all $\bF\in\sF^{m\times n}$, $\br\bx|\widehat\bp|\bF\kt=-i\bnabla\bF(\bx)$, and consequently for $m=3$, we have
	\begin{align}
	&\br\bx|\widehat\bp\cdot|\bF\kt=-i\bnabla\cdot\bF(\bx),
	&&\br\bx|\widehat\bp\times|\bF\kt=-i\bnabla\times \bF(\bx).
	\nn
	\end{align}
With the help of these identities we can express (\ref{Maxwell}) and (\ref{Maxwell-2}) in the form
	\begin{align}
	&\widehat\bp\times|\bcE\kt=k\, \widehat{\hat\bmu}\,|\bcH\kt,
	&&\widehat\bp\times|\bcH\kt=-k\, \widehat{\hat\bvarepsilon}\,|\bcE\kt,
	\label{Maxwell-1}
	\\
	&\widehat\bp\cdot|\bcE\kt=-\widehat\bp\cdot\widehat{\delta\hat\bvarepsilon}\,|\bcE\kt, 
	&&\widehat\bp\cdot|\bcH\kt=-\widehat\bp\cdot\widehat{\delta\hat\bmu}\,|\bcH\kt,
	\label{Maxwell-3}
	\end{align}
where $\widehat{\hat\bmu}:=\hat\bmu(\widehat\bx)$, $\widehat{\hat\bvarepsilon}:=\hat\bvarepsilon(\widehat\bx)$,  $\widehat{\delta\hat\bvarepsilon}:=\delta\hat\bvarepsilon(\widehat\bx)$, $\widehat{\delta\hat\bmu}:=\delta\hat\bmu(\widehat\bx)$, and $\widehat\bx$ is the standard position operator in three dimensions. It is not difficult to see that (\ref{Maxwell-1}) and (\ref{Maxwell-3}) are respectively equivalent to
	\begin{align}
	&\widehat\bQ_-|\bPsi\kt=ik\bSigma_2 \widehat\bfeta|\bPsi\kt,
	\label{Maxwell-4}\\
	&\bp\cdot|\bPsi\kt=-\bp\cdot\widehat\bfeta|\bPsi\kt,
	\label{Maxwell-5}
	\end{align}
where
	\begin{align}
	&\widehat\bQ_\pm:=\bQ_\pm(\widehat\bp),
	&&\bQ_\pm(\bp):=\pm ik\bSigma_2+\bp\times,
	\label{new1}\\
	&\bSigma_2:=\left[\begin{array}{cc}
	\bzero & -i\bI\\
	i\bI & \bzero\end{array}\right]\in\C^{6\times 6},
	&&\widehat\bfeta:=\left[\begin{array}{cc}
	\widehat{\delta\hat\bvarepsilon} &\bzero\\
	\bzero& \widehat{\delta\hat\bmu}\end{array}\right],
	\label{new2}
	\end{align}
and $\bzero$ is the $3\times 3$ zero matrix. 

Next, we use (\ref{Maxwell-5}) and (\ref{new1}) to show that
	\begin{align}
	&\widehat\bQ_+\widehat\bQ_-|\bPsi\kt=-\widehat\bp^{\,2}|\bPsi\kt+k^2 |\bPsi\kt-\widehat\bp\:\widehat\bp\cdot\widehat\bfeta|\bPsi\kt,
	\label{id-1}
	\end{align}
where $\widehat\bp\:\widehat\bp\cdot\widehat\bfeta|\bPsi\kt=\widehat\bp\sum_{i,j=1}^3\widehat p_i\widehat\eta_{ij}|\Psi_j\kt$, and $\widehat\eta_{ij}$ and $|\Psi_j\kt$ are respectively the entries of $\widehat\bfeta$ and $|\bPsi\kt$. Applying $\widehat\bQ_+$ to both sides of (\ref{Maxwell-4}) and making use of (\ref{id-1}), we arrive at the following EM analog of the time-independent Schr\"odinger equation (\ref{sch-eq}).
	\be
	\big(\widehat\bp^{\,2}+\widehat\bV\big)|\bPsi\kt=
	k^2|\bPsi\kt,
	\label{EM-sch-eq}
	\ee
where
	\be
	\widehat\bV:=\big(ik\widehat\bQ_+\bSigma_2+\widehat\bp\,\widehat\bp\cdot\big)\bfeta=\left[\begin{array}{cc}
	(-k^2+\widehat\bp\,\widehat\bp\cdot)\widehat{\delta\hat\bvarepsilon} &
	k\widehat\bp\times \widehat{\delta\hat\bmu}\\
	-k\widehat\bp\times\widehat{\delta\hat\bvarepsilon}&
	(-k^2+\widehat\bp\,\widehat\bp\cdot)\widehat{\delta\hat\bmu} 
	\end{array}\right].
	\label{V=}
	\ee
In the position representation, (\ref{EM-sch-eq}) takes the form 
	\begin{align}
	(\nabla^2+k^2)\br\bx|\bPsi\kt&=(-k^2+k\bSigma_2\,\bnabla\times-\bnabla\,\bnabla\cdot)\bfeta(\bx)\br\bx|\bPsi\kt,
	\label{x-rep}
	\end{align}
 where 
	\[\bfeta(\bx):=\left[\begin{array}{cc}
	\delta\hat\bvarepsilon(\bx) & \bzero\\
	\bzero & \delta\hat\bmu(\bx)\end{array}\right].\]
	
In the following we assume that for all $i,j\in\{1,2,3\}$, $\delta\hat\varepsilon_{ij}(\bx)$, $\delta\hat\mu_{ij}(\bx)$, and their first and second partial derivatives decay to zero faster that $|\bx|^{-2}$ as $|\bx|\to\infty$, so that the solutions of (\ref{x-rep}) tend to a plane wave as $|\bx|\to\infty$. In particular, the outgoing solutions satisfy
	\be
	\br\bx|\bPsi\kt\to\frac{1}{(2\pi)^{3/2}}\left[\bPsi_0\,e^{i\bk\cdot\bx}+\bPhi(\bk',\bk)\,\frac{e^{ik|\bx|}}{|\bx|}\right]
	~~~\for~~~|\bx|\to\infty,
	\label{EM-asymp}
	\ee
where 
	\begin{align}
	&\bPsi_0:=\left[\!\begin{array}{c}
	\bcE_0\\
	\hat\bk\times\bcE_0\end{array}\!\right],
	&&\bPhi(\bk',\bk)=\left[\begin{array}{c}
	\bF(\bk',\bk)\\
	\hat\bx\times\bF(\bk',\bk)\end{array}\right],
	\label{F=}
	\end{align}
$\bcE_0:=\cE_0\,\bfe$, $\cE_0$ is a constant, $\bfe$ is the polarization vector for the incident wave, $\hat\bk:=k^{-1}\,\bk$, $\bk$ is the incident wave vector, $\bk':=k\,\hat\bx$, $\hat\bx:=|\bx|^{-1}\,\bx$, and $\bF$ is an EM analog of the scattering amplitude $f$ that determines the asymptotic values of the amplitude and polarization of the scattered wave. Specifically,
	\be
	\bcE(\bx)\to\frac{1}{(2\pi)^{3/2}}\left[\cE_0\,e^{i\bk\cdot\bx}\,\bfe+\bF(\bk',\bk)\,\frac{e^{ik|\bx|}}{|\bx|}\right]
	~~~\for~~~|\bx|\to\infty,
	\label{EM-asymp_E}
	\ee
which identifies $|\bF(\bk',\bk)|^2$ with the scattering cross section of the medium \cite{TKD}.

The outgoing solutions of (\ref{EM-sch-eq}) fulfill the EM Lippmann-Schwinger equation,
	\be
	|\bPsi\kt=\bPsi_0 |\bk\kt+
	\frac{1}{(2\pi)^{3/2}}\:G(\widehat\bp)\widehat\bV|\bPsi\kt.
	\label{EM-LS-eq}
	\ee
In the momentum representation, this reads
	\begin{align}
     	\br\bp|\bPsi\kt&=\bPsi_0\br\bp |\bk\kt+
	\frac{G(\bp)}{(2\pi)^{3/2}}
	\int_{\mathbb{R}^d} d\bq\, \br\bp|\widehat\bV|\bq\kt 	
	\br\bq|\bPsi\kt.
    	\label{EM-LS-p}
    	\end{align}
According to (\ref{V=}),
	\be
	\br\bp|\widehat\bV|\bq\kt =\big(-k^2+ik\bSigma_2\,\bp\times+\,\bp\,\bp\cdot\big)	\widetilde\bfeta(\bp-\bq),
	\label{tV=}
	\ee
where $\widetilde\bfeta(\bp)$ is the Fourier transform of $\bfeta(\bx)$. 

Similarly to potential scattering, the above-mentioned asymptotic decay condition on $\delta\hat\varepsilon_{ij}(\bx)$ and $\delta\hat\mu_{ij}(\bx)$ implies that $\bPhi(\bk',\bk)$ and consequently the EM scattering amplitude $\bF(\bk',\bk)$ are determined by $\br\bp|\bPsi\kt$ with $|\bp|=k$. It is also easy to see that $\br\bp|\bPsi\kt$ admits the following Born series expansion.
	\be
     	\br\bp|\bPsi\kt=\sum_{n=0}^\infty\br\bp|\bPsi^{(n)}\kt,
    	\label{EM-Born}
    	\ee
where $\br\bp|\bPsi^{(0)}\kt:=\bPsi_0\br\bp|\bk\kt$ and
	\begin{align}
     	\br\bp|\bPsi^{(n+1)}\kt
	&:=\frac{G(\bp)}{(2\pi)^{3/2}}
	\int_{\mathbb{R}^d} d\bq\, \br\bp|\widehat\bV|\bq\kt 	
	\br\bq|\bPsi^{(n)}\kt~~~\for~~~n\geq 0.
    	\label{EM-Born-iteration}
    	\end{align}
In view of (\ref{F=}) and (\ref{EM-Born}), we can also expand the EM scattering amplitude in its Born series according to
	\be
	\bF(\bk',\bk)=\sum_{n=1}^\infty \bF^{(n)}(\bk',\bk),
	\label{Born-F}
	\ee
where $\bF^{(n)}(\bk',\bk)$ are defined by demanding that
	\be
	\br\bx|\bPsi^{(n)}\kt\to\frac{e^{ik|\bx|}}{(2\pi)^{3/2}|\bx|}\left[\begin{array}{c}
	\bF^{(n)}(\bk',\bk)\\
	\hat\bx\times\bF^{(n)}(\bk',\bk)\end{array}\right]~~~\for~~~|\bx|\to\infty~~{\rm and}~~n\geq 1.\nn
	\ee
	
The $N$-th order Born approximation amounts to neglecting the terms $\br\bp|\bPsi^{(n)}\kt$ and $\bF^{(n)}(\bk',\bk)$ with $n>N$ in (\ref{EM-Born}) and (\ref{Born-F}), respectively. It gives the exact solution of the EM scattering problem if
	\be
	\br\bp|\bPsi\kt=\sum_{n=0}^{N}\br\bp|\bPsi^{(n)}\kt~~~~\for~~~~|\bp|=k.
	\label{EM-exact}
	\ee
This in turn implies $\bF(\bk',\bk)=\sum_{n=1}^N \bF^{(n)}(\bk',\bk)$.

Next, consider situations where
	\begin{align}
	\widetilde{\bfeta}(\bp)=\bzero~~~\for~~~\bu\cdot\bp<\alpha,
	\nn
	\end{align}
for some $\alpha\in\R^+$ and $\bu\in\R^3$ such that $|\bu|=1$, i.e.,
for all $i,j\in\{1,2,3\}$,
	\begin{align}
	\widetilde{\delta\hat\varepsilon}_{ij}(\bp)=
	\widetilde{\delta\hat\mu}_{ij}(\bp)=0~~~\for~~~\bu\cdot\bp<\alpha.
	\label{EM-potential}
	\end{align}
Then, (\ref{tV=}) implies 
	\be
	\br\bp|\widehat\bV|\bq\kt=\bzero~~~\for~~~\bu\cdot(\bp-\bq)<\alpha.
	\nn
	\ee
In view of this relation and (\ref{EM-LS-p}) and (\ref{EM-Born-iteration}), we can easily repeat the analysis of Sec.~2 and arrive at the following EM analog of Theorem~1.
	\begin{itemize}
	\item[]{\bf Theorem 2}: Consider the scattering of an incident time-harmonic plane EM wave with wavenumber $k$ by a stationary linear medium with relative permittivity and permeability tensors, $\hat\bvarepsilon(\bx)$ and  $\hat\bmu(\bx)$. Let $\delta\hat\bvarepsilon(\bx):=\hat\bvarepsilon(\bx) -\bI$, $\delta\hat\bmu(\bx):=\hat\bmu(\bx) -\bI$, $\alpha\in\R^+$, $N:=\lfloor 2k/\alpha\rfloor$, and $\bu\in\R^3$ such that $|\bu|=1$. Suppose that the entries of $\delta\hat\bvarepsilon(\bx)$ and $\delta\hat\bmu(\bx)$ and their first and second partial derivatives decay to zero faster than $|\bx|^{-2}$ as $|\bx|\to\infty$, and their Fourier transforms satisfy (\ref{EM-potential}). Then the $N$-th order Born approximation provides the exact solution of the scattering problem. In particular, the medium does not scatter incident waves with $k<\alpha/2$, and the first Born approximation is exact for $k<\alpha$.
	\end{itemize}

\section{Concluding Remarks} 

Among the basic issues one must address in the study of any consistent approximation scheme is the search for conditions under which it produces the exact solution of the problem. Typical examples in quantum mechanics are the adiabatic and semiclassical approximations. Adiabatic approximation gives the exact expression for the evolution operator of a time-dependent Hamiltonian $\widehat H(t)$ with a discrete spectrum, if and only if $\widehat H(t)$ has a complete set of time-independent eigenvectors. Semiclassical approximation gives the exact expression for the eigenfunctions of a standard Hamiltonian in one dimension, if the potential is piecewise constant. Unlike adiabatic and semiclassical approximations, the task of finding conditions for the exactness of the Born approximation has eluded several generations of physicists. Recently, we used the dynamical formulation of stationary scattering  of Refs.~\cite{pra-2021,pra-2023a} to identify scattering problems in two and three dimensions that are exactly solvable by the first-order Born approximation \cite{pra-2021,ptep-2024}. In the case of EM scattering, this relied on the technical assumption pertaining the existence of a coordinate system where the detectors are placed on the planes $x_3=\pm\infty$, and $\hat\varepsilon_{33}(\bx)$ and $\hat\mu_{33}(\bx)$ do not vanish, \cite{ptep-2024}. 

In the present article, we use the well-known formulation of stationary scattering based on the Lippmann-Schwinger equation to extend the results of Refs.~\cite{pra-2021,ptep-2024} on the exactness of the first-order Born approximation for incident waves whose wavenumber $k$ does not exceed a critical value $\alpha$. Specifically, we show that for $k>\alpha$ the first-order Born approximation fails, but there is a positive integer $N$ such that the $N$-th order Born approximation gives the exact solution of the scattering problem. The approach we pursue here has the advantage of not requiring the knowledge of the dynamical formulation of stationary scattering. In particular, in the case of EM scattering, it also applies to situations where $\hat\varepsilon_{33}(\bx)$ or $\hat\mu_{33}(\bx)$ can be zero for some $\bx\in\R^3$.

It is easy to see that the conditions of Theorems 1 and 2 require the corresponding interaction potentials and the permittivity-permeability profiles to be respectively complex-valued and complex matrix-valued functions \cite{apl-2023}. This shows that the acoustic and optical realizations of these systems must involve carefully engineered regions of gain and loss. Recent experimental studies \cite{jiang-2017,ye-2017} of unidirectional invisibility in optical media fulfilling spatial Kramers-Kronig relations \cite{horsley-2015,longhi-2015} show that this kind of technical difficulties can be overcome in one dimension.  \vspace{6pt}

\noindent{\bf Acknowledgements}:
This work has been supported by Turkish Academy of Sciences (T\"UBA).

\ed